\documentclass{iau}
\usepackage{graphicx}
\usepackage[utf8]{inputenc}
\usepackage{amsmath}
\usepackage{natbib}
\usepackage{aas_macros}

\title[Making Kepler 11] %% give here short title %%
{Making systems of Super Earths by inward migration of planetary embryos}

\author[Christophe Cossou, Sean N. Raymond \and Arnaud Pierens]   %% give here short author
{Christophe Cossou$^{1,2}$ \and Sean N. Raymond$^{1,2}$ \and Arnaud Pierens$^{1,2}$}

\affiliation{$^1$Univ. Bordeaux, LAB, UMR 5804, F-33270, Floirac, France.\\[\affilskip]
$^2$CNRS, LAB, UMR 5804, F-33270, Floirac, France}

\pubyear{2013}
\volume{299}  %% insert here IAU Symposium No.
\pagerange{119--126}
\jname{Title of your IAU Symposium}
\editors{A.C. Editor, B.D. Editor \& C.E. Editor, eds.}

\newcommand{\unit}[1]{\ensuremath{\mathrm{\thinspace #1}}}%allow to write properly physical units such as Hz, in math or text mode.
\newcommand{\mearth}{\unit{M_\oplus}}
\newcommand{\od}[2]{\frac{\mathrm{d}#1}{\mathrm{d}#2}}
\newcommand{\reffig}[1]{\figurename~\ref{#1}}
\newcommand{\MMR}[2]{\mbox{#1:#2}}
\newcommand{\inv}[1]{\frac{1}{#1}}

\begin{document}

\maketitle

\begin{abstract}
Using N-body simulations with planet-disk interactions, we present a mechanism capable of forming compact systems of hot super Earths such as Kepler 11. Recent studies show that outward migration is common in the inner parts of radiative disks. However we show that two processes naturally tip the balance in favor of inward migration. First the corotation torque is too weak to generate outward migration for planetary embryos less massive than $4\mearth$. Second, system of multiple embryos generate sustained non-zero eccentricities that damp the corotation torque and again favor inward migration. Migration and accretion of planetary embryos in realistic disks naturally produce super Earths in resonant chains near the disk inner edge. Their compact configuration is similar to the observed systems.
\end{abstract}

\firstsection % if your document starts with a section,
\section{Introduction}%TODO ref for super earth are common !!!!
Super Earths ($1-10\mearth$) are common \citep{mayor2011road, howard2012occurrence} even if none exist in our own solar system.  They tend to be found in systems of multiple Super Earths in close-in, compact orbital configurations.  For example, the Kepler 11 system contains five planets within the orbit of Mercury, each more massive than Earth \citep{lissauer2011closely}. 

Here we show how that gas disk-driven orbital migration naturally produces systems similar to the observed ones.  We use N-body simulations that include torques from a 1-D gaseous disk to show how a system of planetary embryos starting at several AU embedded naturally migrates inward and accretes into a compact system of hot Super Earths.  Some embryos can also grow and remain trapped on more distant orbits and presumably become giant planet cores.

\section{Methods}\label{sec:model}
We start from a 1D protoplanetary disk model, with parameters listed in Table~\ref{tab:disk_param}. The key parameter is the surface density profile, defined as a power law in radial range $0.1<R<100\unit{AU}$ : 
\begin{align}
\Sigma(R) &= \Sigma_0 \left(\frac{R}{1\unit{AU}}\right)^{-d}
\end{align}
The inner edge of the disk is smoothed with a $\tanh$ function on a length scale of $H(R_\text{in})$, where $H$ is the scaleheight of the disk, so the density goes to zero at the inner edge.  From the surface density profile, we self-consistently compute the temperature $T$, thermal diffusivity $\chi$, scaleheight $H$ and optical depth profiles $\tau$. To calculate the temperature we use the following energy equation : 
\begin{align}
0 &= - C_\text{BB} + H_\text{en} + H_\text{irr} + H_\text{vis}
\end{align}\label{eq:energy_equation}
with
\begin{align*}
C_\text{BB} &= 2\sigma \frac{T^4}{\frac{3}{8}\tau + \frac{\sqrt{3}}{4} + \inv{4\tau}} & H_\text{en} &= 2 \sigma {T_\text{en}}^4 \\
H_\text{irr} &= 2 \sigma {T_\star}^4 \frac{{R_\star}^2}{r^2} (1-\varepsilon) * \left[0.4 \frac{R_\star}{r} + r \od{}{r}\left(\frac{H}{r}\right)\right] & H_\text{vis} &= \frac{9}{4} \nu\Sigma\Omega^2
\end{align*}
$C_\text{BB}$ is the disk's black body cooling and $H_\text{en}$ is the black body heating by the disk envelope. $H_\text{vis}$ is the viscous heating. $H_\text{irr}$ is the heating from stellar irradiation~\citep{chiang1997spectral, menou2004low}, $\sigma$ the Stephan-Boltzmann constant, $\nu$ the viscosity of the disk, $T_\star$ and $R_\star$ the temperature and radius of the central star respectively, $\Omega$ the angular speed in the disk at a given position and $\varepsilon$ the disk's albedo. Starting at the outer edge of the disk, where we impose that $T=10\unit{K}$, we retrieve the temperature profile by solving eq. (\ref{eq:energy_equation}) numerically.

We use the formulae of \cite{paardekooper2011torque} to compute the torque exerted by our 1D disk on a planet of a given mass at a given orbital radius.  The main differences with the model described in this paper are : 
\begin{itemize}
\item The temperature profile is not a power law in our model, but instead a local power law between each point of a table of several hundred points spaced in orbital distance
\item the scaleheight of the disk, defined as $H = \frac{1}{\Omega}\sqrt{\frac{k_B T}{\mu m_H}}$, is computed following the temperature profile, with $k_B$ the Boltzmann constant, $\mu$ the mean molecular weight and $m_H$ the mass of an hydrogen atom.
\end{itemize}
We implement type I eccentricity and inclination damping following \cite{cresswell2008three}.  We also include an eccentricity-migration feedback whereby the corotation torque is weakened for eccentric orbits \citep{bitsch2010orbital}, using the equations from \citep{cossou2013convergence}.

\begin{table}[htb]
\centering
\begin{tabular}{|c|c|c|c|c|c|}
\hline
$b/h = 0.6$ & $\gamma = 7/5$ & $\mu = 2.35$ & $\alpha = 5\cdot 10^{-3}$ & $T_\star = 5700\unit{K}$ & $R_\star = 4.65\cdot 10^{-3}\unit{AU}$\\\hline
\multicolumn{2}{|c|}{$\text{Disk Albedo} = 0.5$} & \multicolumn{2}{|c|}{$\mathrm{Disk : R\in[0.1;100]\unit{AU}}$} & \multicolumn{2}{|c|}{$\Sigma(R) = 300 \cdot R^{-1/2}\unit{g/cm^2}$}\\\hline
\end{tabular}
\caption{Parameters of the disk. In addition to thoses parameters, note that the opacities were retrieved from the opacity table of \cite{hure2000transition}. The viscosity is modeled via the $\alpha$-prescription \citep{shakura1973black}.}\label{tab:disk_param}
\end{table}

Disk forces were added to the hybrid version of the {\tt Mercury} integrator \citep{chambers1999hybrid}.  Drag and migration forces are not applied to objects inside the inner cavity (inside 0.1 AU).  Collisions were treated as inelastic mergers. 

\section{Mechanism}
\begin{figure}[htb]
\centering
\includegraphics[width=0.48\linewidth]{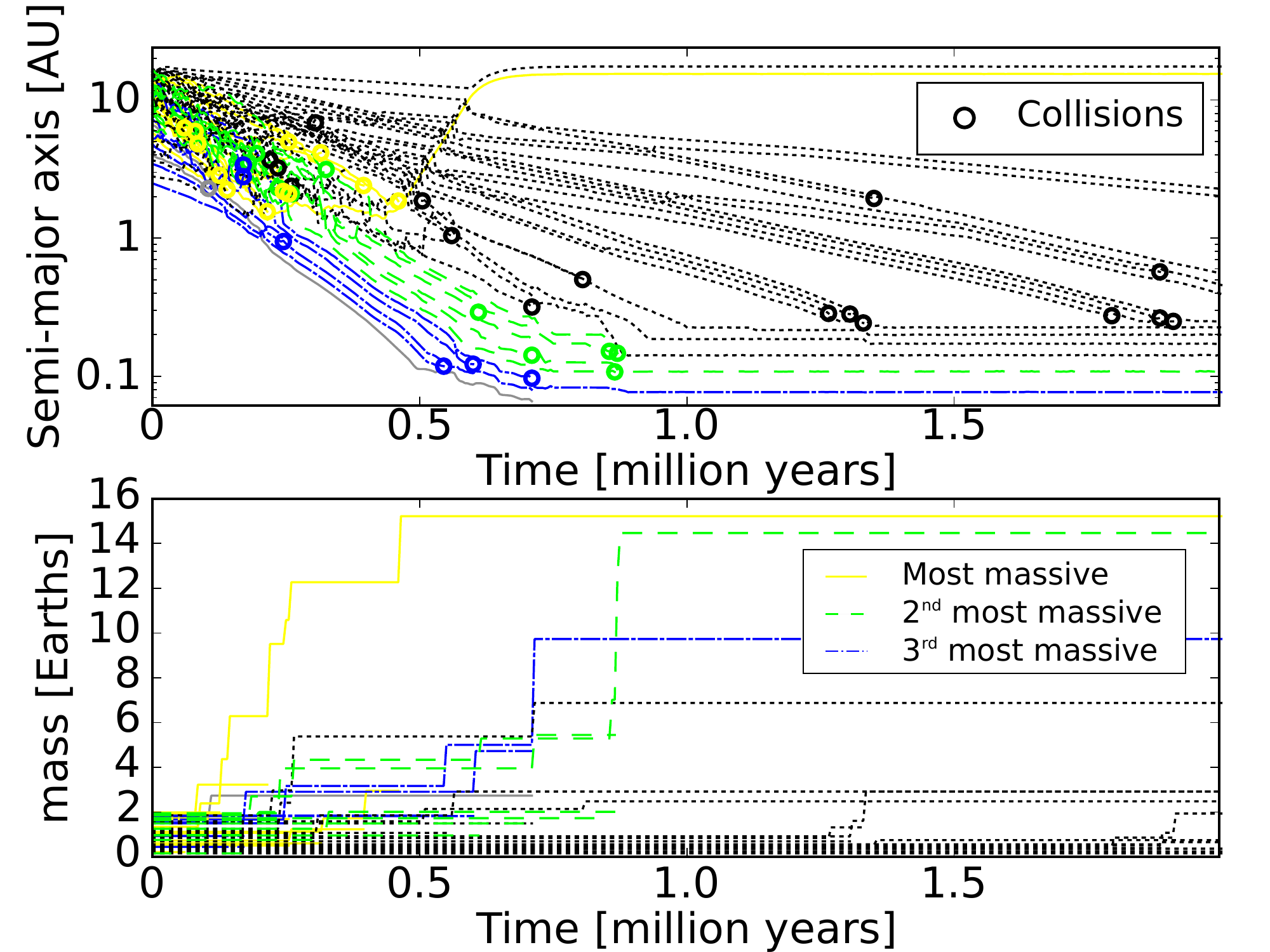}\hfill\includegraphics[width=0.48\linewidth]{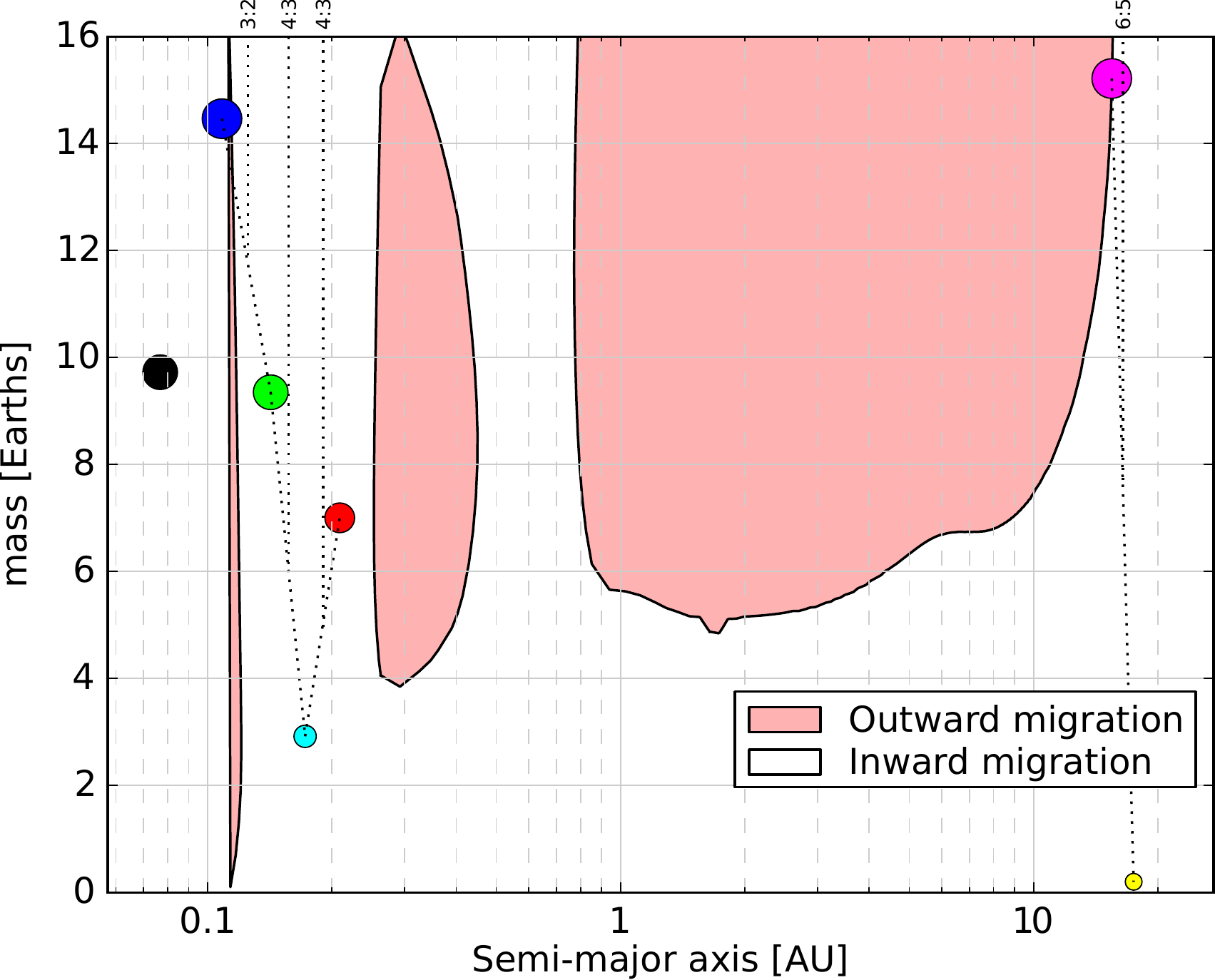}
\caption{A simulation that forms a compact system of hot Super Earths. \textbf{Left}: The orbital evolution (top) and mass growth (bottom) of the embryos. \textbf{Right}: The final state of the system. Black lines represent zero-torque zones where isolated planets should stop migrating.}\label{fig:simulation}
\end{figure}

The right panel of \reffig{fig:simulation} shows the disk's migration map. Two zones of convergent migration exist. The first convergence zone causes all embryos between $0.2$ and $0.8\unit{AU}$ to migrate toward $0.5\unit{AU}$.  The second causes embryos between $0.9$ and $\sim 100\unit{AU}$ to migrate toward $\sim15\unit{AU}$. However, multiple planet systems do not actually migrate to convergence zone like isolated planets. Rather, they become trapped in chains of mean motion resonances.  The resonant configurations sustain eccentricities large enough to attenuate the corotation torque \citep{bitsch2010orbital}. The system stabilizes in an equilibrium position in the disk where the sum of all the torques felt by the planets cancels out \citep{cossou2013convergence}.

The left panel of \reffig{fig:simulation} shows the dynamics and accretion of a system of embryos embedded in the disk.  The embryos comprised $60\mearth$ in total, and started with masses randomly chosen from $0.1-2 \mearth$ spaced from $2-17\unit{AU}$, and the integration lasted for 10 Myr.  At early times, embryos migrate inward because their masses are below the $\sim 4 \mearth$ threshold for outward migration (see right panel of \reffig{fig:simulation}). Because of their different masses the embryos do not all migrate at the same rate. This leads to close encounters and collisions.  Almost all of the embryos migrate inward and complete their growth close to the inner edge of the disk.  The final configuration of planets is a compact resonant system.  The most massive planet grows fast enough to reverse its migration and stabilizes at $15\unit{AU}$.  While migrating outward, it traps a lower mass planet in \MMR{6}{5} resonance.  The low-mass planet is pushed by the more massive one, and the two-planet system is effectively  ruled by the zero-migration line of the more massive planet.

\begin{figure}[htb]
\centering
\includegraphics[width=0.9\linewidth]{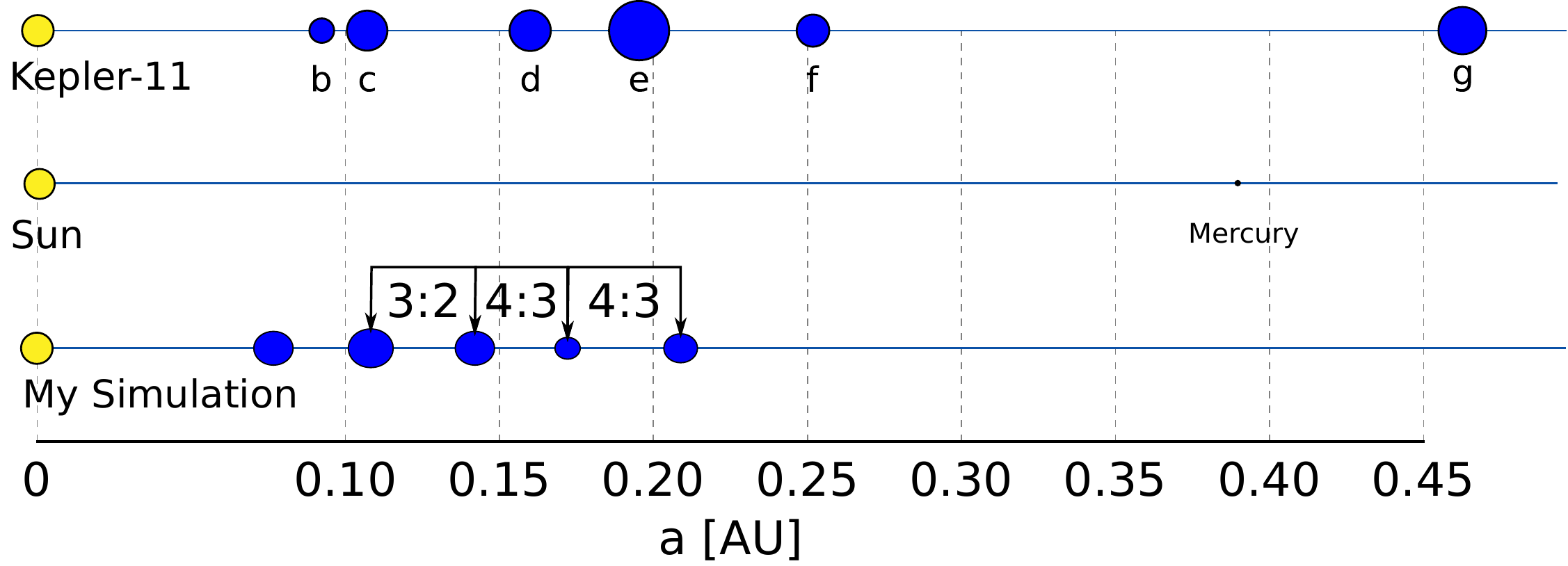}
\caption{Final orbital configuration of the inner part of the simulation compared with the Solar System and Kepler 11.}\label{fig:kepler-comparison}
\end{figure}

The embryos ended up at the inner edge of the disk for two reasons.  First, objects less massive than $\sim 4 \mearth$ simply cannot migrate outward (see \reffig{fig:simulation}).  During its inward migration, an embryo must accrete quickly if it is to enter a zone of outward migration, as was the case for the outer massive planet.  Second,  embryos that do migrate outward may encounter other large embryos and become trapped in resonance, leading to sustained non-zero eccentricities that damp the corotation torque \citep{bitsch2010orbital} and tip the balance toward inward migration \citep{cossou2013convergence}.  In this simulation the first mechanism dominated but both mechanisms can be important.  The system was held up because a huge positive torque exists close to the inner edge due to the sudden decrease of the surface density profile \citep{masset2006disk}.  Our results are thus similar to those of \cite{terquem2007migration} even though outward migration does occur in our modeled disk.  

\reffig{fig:kepler-comparison} shows the inner simulated system compared to our Solar System and the Kepler-11 system.  The simulated system has similar masses to Kepler-11 but its orbital configuration is even more compact, although there are no planets in the simulation between 0.25 and 15 AU.  The four outer simulated planets are in a resonant chain but the innermost planet is not.  This is because the innermost planet was pushed into the disk's inner gas-free cavity where it lacked the energy dissipation required for efficient resonant trapping.

\section{Conclusion}

In our disk model, a compact system of hot Super Earths is formed by migration and accretion of planetary embryos.  Inward migration is favored.  Low-mass embryos naturally migrate inward.  Outward migration is stalled or even reversed by the eccentricities sustained through resonances~\citep{cossou2013convergence}.  In essence, the presence of multiple large embryos leads to non-zero eccentricities such that they no longer follow the migration map (right panel of \reffig{fig:simulation}).  

This formation mechanism for hot Super Earths is robust against a much wider range of disk parameters than the one presented here.  We note that the survival of the simulated hot Super Earths may depend on the detailed structure of the inner edge of the gas disk, which provides the positive torque needed to balance the inward-driven system.  Their survival also likely depends on how the disk evolves \citep{horn2012orbital} and especially on how it dissipates.  We plan to take these effects, and others such as stochastic forcing from turbulence, into account in future work.  

\bibliographystyle{apj}
\bibliography{biblio}

\end{document}